\documentclass[letter,10point,onecolumn,preprint]{emulateapj}

\usepackage{graphics,graphicx,pdflscape,subfigure,apjfonts}

\def\cxo{{\em Chandra}}
\def\xmm{{\em XMM}}
\def\asca{{\em ASCA}}
\def\chic{ChIcAGO}
\def\erg{erg~cm$^{-2}$~s$^{-1}$}
\def\etal{et al.}

\begin{document}

\title{Identification of a Population of X-ray Emitting Massive Stars in the Galactic Plane}
\shorttitle{A New Population of X-ray Emitting Massive Stars}
\shortauthors{Anderson et al.}
\submitted{Accepted to {\em The Astrophysical Journal}}
\author{Gemma E. Anderson,\altaffilmark{1} B. M. Gaensler,\altaffilmark{1,$\dagger$} David L. Kaplan,\altaffilmark{2, $\ddagger$} Bettina Posselt,\altaffilmark{3} Patrick O. Slane,\altaffilmark{3}\\
Stephen S. Murray,\altaffilmark{3} Jon C. Mauerhan,\altaffilmark{4} Robert A. Benjamin,\altaffilmark{5} Crystal L. Brogan,\altaffilmark{6} Deepto Chakrabarty,\altaffilmark{7}\\ Jeremy J. Drake,\altaffilmark{3} Janet E. Drew,\altaffilmark{8} Jonathan E. Grindlay,\altaffilmark{3} Jaesub Hong,\altaffilmark{3} T. Joseph W. Lazio,\altaffilmark{9} Julia C. Lee,\altaffilmark{3}\\
 Danny T. H. Steeghs,\altaffilmark{10} Marten H. van Kerkwijk,\altaffilmark{11}}
\altaffiltext{1}{Sydney Institute for Astronomy, School of Physics A29, The University of Sydney, NSW 2006, Australia: g.anderson@physics.usyd.edu.au}
\altaffiltext{2}{Hubble Fellow; Kavli Institute for Theoretical Physics, Kohn Hall, University of California, Santa Barbara, CA 93106, USA}
\altaffiltext{3}{Harvard-Smithsonian Center for Astrophysics, Cambridge, MA 02138, USA}
\altaffiltext{4}{Spitzer Science Center, California Institute of Technology, Pasadena, CA 91125, USA}
\altaffiltext{5}{Department of Physics, University of Wisconsin, Whitewater, WI 53190, USA}
\altaffiltext{6}{National Radio Astronomy Observatory, Charlottesville, VA 22903, USA}
\altaffiltext{7}{MIT Kavli Institute for Astrophysics and Space Research and Department of Physics, Massachusetts Institute of Technology, Cambridge, MA 02139, USA}
\altaffiltext{8}{Centre for Astrophysics Research, STRI, University of Hertfordshire, Hatfield AL10 9AB, UK}
\altaffiltext{9}{Naval Research Laboratory, Washington, DC 20375, USA}
\altaffiltext{10}{Department of Physics, University of Warwick, Coventry CV4 7AL, UK}
\altaffiltext{11}{Department of Astronomy and Astrophysics, University of Toronto, Toronto, ON M5S 3H4, Canada}
\altaffiltext{$\dagger$}{Australian Research Council Federation Fellow}
\altaffiltext{$\ddagger$}{Current address: Department of Physics, University of Wisconsin, Milwaukee, WI 53201-0431, USA}

\begin{abstract}

We present X-ray, infrared, optical and radio observations of four previously unidentified Galactic plane X-ray sources, AX~J163252--4746, AX~J184738--0156, AX~J144701--5919 and AX~J144547--5931. Detection of each source with the \textit{Chandra X-ray Observatory} has provided sub-arcsecond localizations, which we use to identify bright infrared counterparts to all four objects. Infrared and optical spectroscopy of these counterparts demonstrate that all four X-ray sources are extremely massive stars, with spectral classifications Ofpe/WN9 (AX~J163252--4746), WN7 (AX~J184738--0156 = WR121a), WN7--8h (AX~J144701--5919) and OIf$^+$ (AX~J144547--5931). AX~J163252--4746 and AX~J184738--0156 are both luminous, hard, X-ray emitters with strong Fe {\sc xxv} emission lines in their X-ray spectra at $\sim6.7$~keV. The multi-wavelength properties of AX~J163252--4746 and AX~J184738--0156 are not consistent with isolated massive stars or accretion onto a compact companion; we conclude that their X-ray emission is most likely generated in a colliding-wind binary system. For both AX~J144701--5919 and AX~J144547--5931, the X-ray emission is an order of magnitude less luminous and with a softer spectrum. These properties are consistent with a colliding-wind binary interpretation for these two sources also, but other mechanisms for the generation of X-rays cannot be excluded. There are many other as yet unidentified X-ray sources in the Galactic plane, with X-ray properties similar to those seen for AX~J163252--4746, AX~J184738--0156, AX~J144701--5919 and AX~J144547--5931. This may indicate a substantial population of X-ray-emitting massive stars and colliding-wind binaries in the Milky Way.

\end{abstract}


\section{Introduction}

Wolf-Rayet (WR) stars and their O-type supergiant progenitors (Of) evolve from the most massive stars in our Galaxy, with initial masses $\gtrsim25 M_{\odot}$. These evolved stars, particularly WR, have extremely strong stellar winds, experiencing high mass-loss rates of $\dot{M} \sim 10^{-5} M_{\odot}$yr$^{-1}$ and, in some cases, have luminosities $>10^{6} L_{\odot}$ \citep{crowther08}. Their short lifetimes make them very rare; $<400$ WR stars are known in our Galaxy \citep{vanderhucht06,martins08,shara09,mauerhan09,mauerhan10}, and are usually only found in the Galactic plane.

Massive stars have historically been discovered through optical and infrared observations and are classified based on their spectral characteristics in these wavebands. However, X-ray observations are now becoming a newly recognized technique for discovering massive stars and are also a powerful tool in assessing their physical environments \citep[e.g.][]{mauerhan10}, allowing us to determine if they are isolated or in a high-mass X-ray binary (HMXB) or colliding-wind binary (CWB) system. By discovering more of these massive stars and understanding their emission mechanisms, we can determine how mass-loss drives the different stages of stellar evolution.

The most accepted model for X-ray generation in a single hot star is the instability-driven wind-shock picture, which attributes the production of soft, thermal X-ray emission, with temperatures of $kT <1$ keV, to shocks distributed throughout the wind \citep{lucy80,lucy82}. More exotic models of X-ray generation in isolated massive stars include the magnetically channeled wind-shock mechanism, which was first explored in detail by \citet{babel97b,babel97a}. In this case a radiatively driven stellar wind is magnetically channeled from the two hemispheres of the star. These streams collide at the magnetic equator and the rapid deceleration causes shock heating resulting in high levels of hard thermal X-ray emission. There is also the possibility of isolated OB supergiant stars producing intrinsic hard non-thermal X-rays through inverse Compton scattering \citep{chen91}. 

Alternatively, the X-ray emission may not be completely intrinsic to the massive star but created through a binary interaction. X-rays can be generated in high-mass X-ray binaries (HMXBs) through gravitational accretion onto a compact object such as a neutron star (NS) or black hole (BH). The two main classes of HMXBs are the Be X-ray binary systems (BeX) and the supergiant X-ray binaries (SGXB). BeXs are accretion fed by a disk around a Be star and are often transient X-ray sources whereas SGXBs are wind-fed, and persistent, X-ray sources \citep{mcclintock06}.

Colliding-wind binary (CWB) systems are another class of massive stellar binaries marked by extreme wind loss and high-energy emission. Originally predicted by \citet{prilutskii76} and \citet{cherepashchuk76}, the supersonic winds from the two massive stars in a binary produce shock-heated gas \citep{stevens92}, resulting in hard, thermal X-ray emission \citep{pittard10} and possibly $\gamma$-rays, likely produced by inverse Compton scattering \citep[e.g.][]{benaglia03,pittard06,reimer06,debecker07}. The detection of this high-energy emission allows us to probe the nature of these shocks and provides a laboratory for investigating particle acceleration in a very different density regime to that in supernova remnants.

In this Paper we discuss our classification of four previously unidentified Galactic X-ray sources; \object[AX J1632.8-4746]{AX J163252--4746}, \object[AX J1847.6-0156]{AX J184738--0156}, \object[AX J1447.0-5919]{AX J144701--5919} and \object[AX J1445.7-5931]{AX J144547--5931}. These sources have been observed with the \textit{Chandra X-ray Observatory} as part of the ``ChIcAGO'' ({\em \underline{Ch}asing the \underline{I}dentifi\underline{c}ation of \underline{A}SCA \underline{G}alactic \underline{O}bjects}) project (Anderson \etal\ in prep.), a survey designed to localize and classify the unidentified X-ray sources discovered during the \asca\ Galactic Plane Survey \citep[AGPS;][]{sugizaki01}. \asca\ (the Advanced Satellite for Cosmology and Astrophysics) surveyed the inner region of the Galactic plane, detecting 163 X-ray sources with fluxes between $10^{-11}$ and $10^{-13}$ \erg\ in the $0.7-10.0$ keV energy range. Due to the \asca\ X-ray telescope's $\sim3'$ spatial resolution, only a third of these X-ray sources have been properly characterized and little is known about the remaining unidentified objects. This unidentified population should contain at least some rare classes of X-ray objects, as modeling by \citet{hands04} and \citet{grindlay05} has demonstrated that the Galactic populations of cataclysmic variables, bright X-ray binaries, and background Active Galactic Nuclei cannot account for the entire observed flux distribution of X-ray sources detected in the AGPS. In ChIcAGO, we are combining the sub-arcsecond localization capabilities of \cxo\ with a detailed multi-wavelength follow-up program, with the goal of classifying the $> 100$ unidentified sources in the AGPS. 

We chose to begin our investigation with AX J163252--4746, AX J184738--0156, AX J144701--5919 and AX J144547--5931 (listed in order of decreasing X-ray flux) as they are all highly absorbed with bright infrared counterparts (8~{\micron} magnitude $<7$), making them distant, luminous and unusual objects. The \cxo\ observations and follow-up studies are described in $\S 2$, along with the results from these observations. In $\S 3$ we discuss the X-ray spectra of these sources and classification of their counterparts, demonstrating that these four objects are all X-ray emitting massive stars.

\section{Observations and Results}

\subsection{\cxo\ and Archival \textit{XMM Newton} Data}

Our \cxo\ observations of AX J163252--4746, AX J184738--0156, AX J144701--5919 and AX J144547--5931 were short (between $1-3$ ks per source). This allowed us to detect $\sim100$ counts per target so as to fulfill the primary aim to localize each source, while providing limited spectral information. We observed AX J184738--0156, AX J144701--5919 and AX J144547--5931 with the Advanced CCD Imaging Spectrometer \citep[ACIS;][]{garmire03} in the ACIS-S mode. As AX J163252--4746 has a predicted ACIS count-rate $>0.2$ counts s$^{-1}$, we used the High Resolution Camera \citep[HRC;][]{murray00} in the HRC-I mode in order to avoid the positional degradation associated with pile-up \citep{davis01} that would arise from an ACIS observation.

\begin{figure}
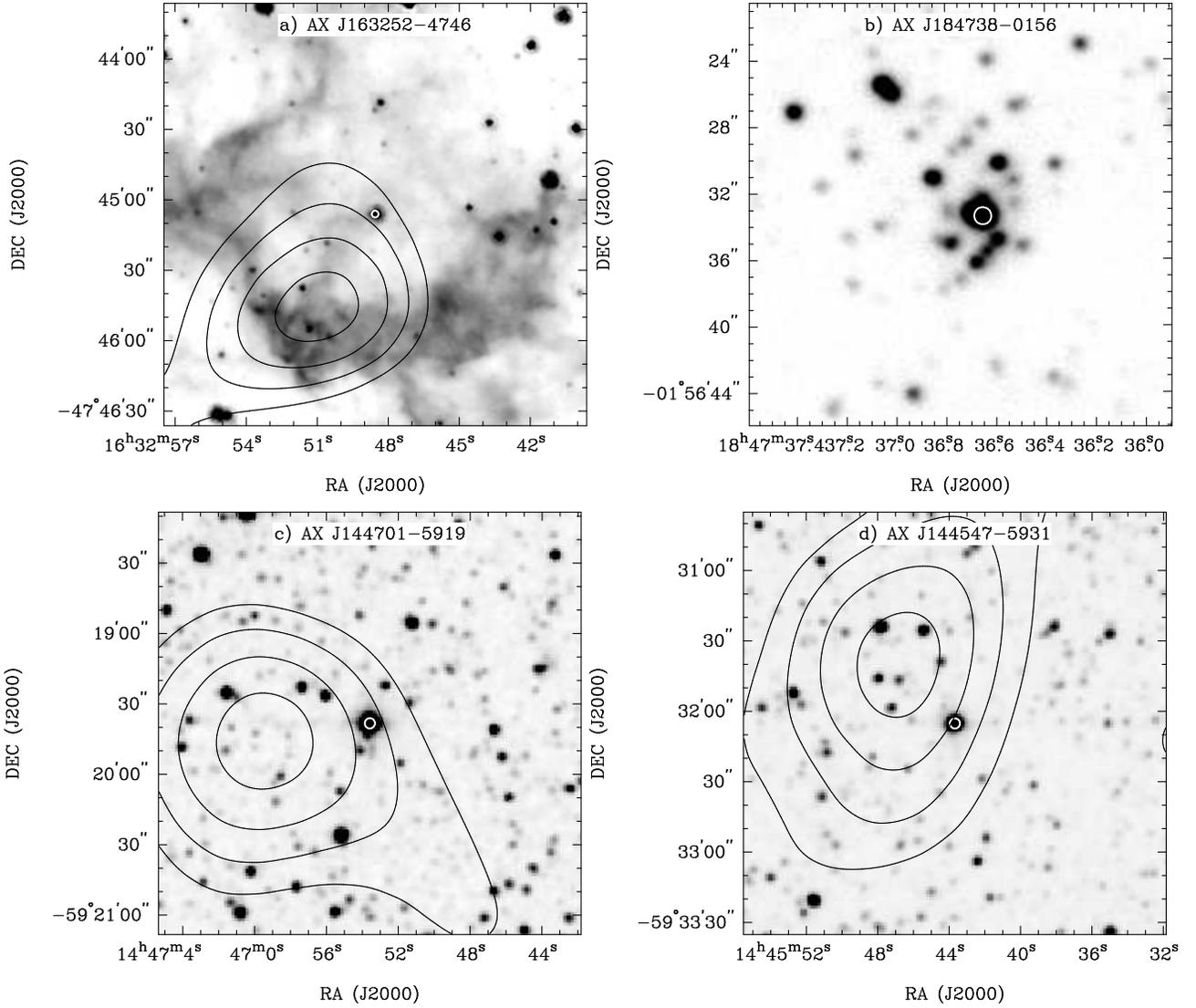

\begin{center}
\subfigure{\label{f1a}\includegraphics[width=0.4\textwidth, angle=270]{f1a.ps}}
\subfigure{\label{f1b}\includegraphics[width=0.4\textwidth, angle=270]{f1b.ps}}
\subfigure{\label{f1c}\includegraphics[width=0.4\textwidth, angle=270]{f1c.ps}}
\subfigure{\label{f1d}\includegraphics[width=0.4\textwidth, angle=270]{f1d.ps}}
\caption{Infrared images of the regions surrounding AX J163252--4746, AX J184738--0156, AX J144701--5919 and AX J144547--5931. The white contours represent the smoothed \cxo\ detection of AX J163252--4746 and AX J184738--0156 (each at 95\% of the peak count-rate), and AX J144701--5919  and AX J144547--5931 (each at 40\% of the peak count-rate). In each case the infrared counterpart to the X-ray source is clearly detected. The black contours show the original \asca\ detection, in the $0.7-10.0$ keV energy range, of each source at 65, 75, 85 and 95\% of the peak count-rate. a) GLIMPSE 8{\micron} image of AX J163252--4746. b) PANIC $H$-band image of the star cluster in which AX J184738--0156 is embedded. The \cxo\ contour is clearly coincident with the brightest member of the cluster. The field of view of this image is too small to overlay the \asca\ contours. c) 2MASS $K$-band image of AX J144701--5919. d) 2MASS $K$-band image of AX J144547--5919.}
\label{fig1}
\end{center}
\end{figure}

These data were reduced using the Chandra Interactive Analysis of Observation software (CIAO), version 4.1, following the standard reduction recipes given in the online CIAO 4.1 Science Threads\footnote{http://cxc.harvard.edu/ciao/threads/}. Further details on the \cxo\ observations will be published by Anderson et al. (in prep). For each of the four AGPS targets, a \cxo\ source with $50-150$ counts was detected within $3'$ of the published \asca\ position. The need for \cxo\ observations to localize the AGPS sources is illustrated in Figure~\ref{fig1}, where each \cxo\ detection clearly shows a much more precise source position (white contour) when compared to the original \asca\ detection (black contours). The \cxo\ instruments and exposure times used for the observations can be found in Table~\ref{tab:1}. The resulting \cxo\ position for each source is listed in Table~\ref{tab:2}. 

AX J163252--4746 and AX J184738--0156 were also detected off-axis in archival \textit{XMM-Newton} data (observational ID's 0201700301 and 0203850101, respectively), with $>9600$ counts each. These two sources are catalogued under the names 2XMM J163248.6--474504 and 2XMM J184736.6--015633, respectively, in the \textit{XMM-Newton} Serendipitous Source catalogue \citep[2XMMi:][]{watson09}. The details on these observations can also be found in Table~\ref{tab:1}. (AX J163252--4746 was also detected off-axis in ten other \xmm\ observations. However, we chose to concentrate our analysis on observational ID 0201700301 as it has the longest integration and therefore the greatest number of counts detected.)

\begin{figure}
\begin{center}
\subfigure{\label{f2a}\includegraphics[width=0.6\textwidth]{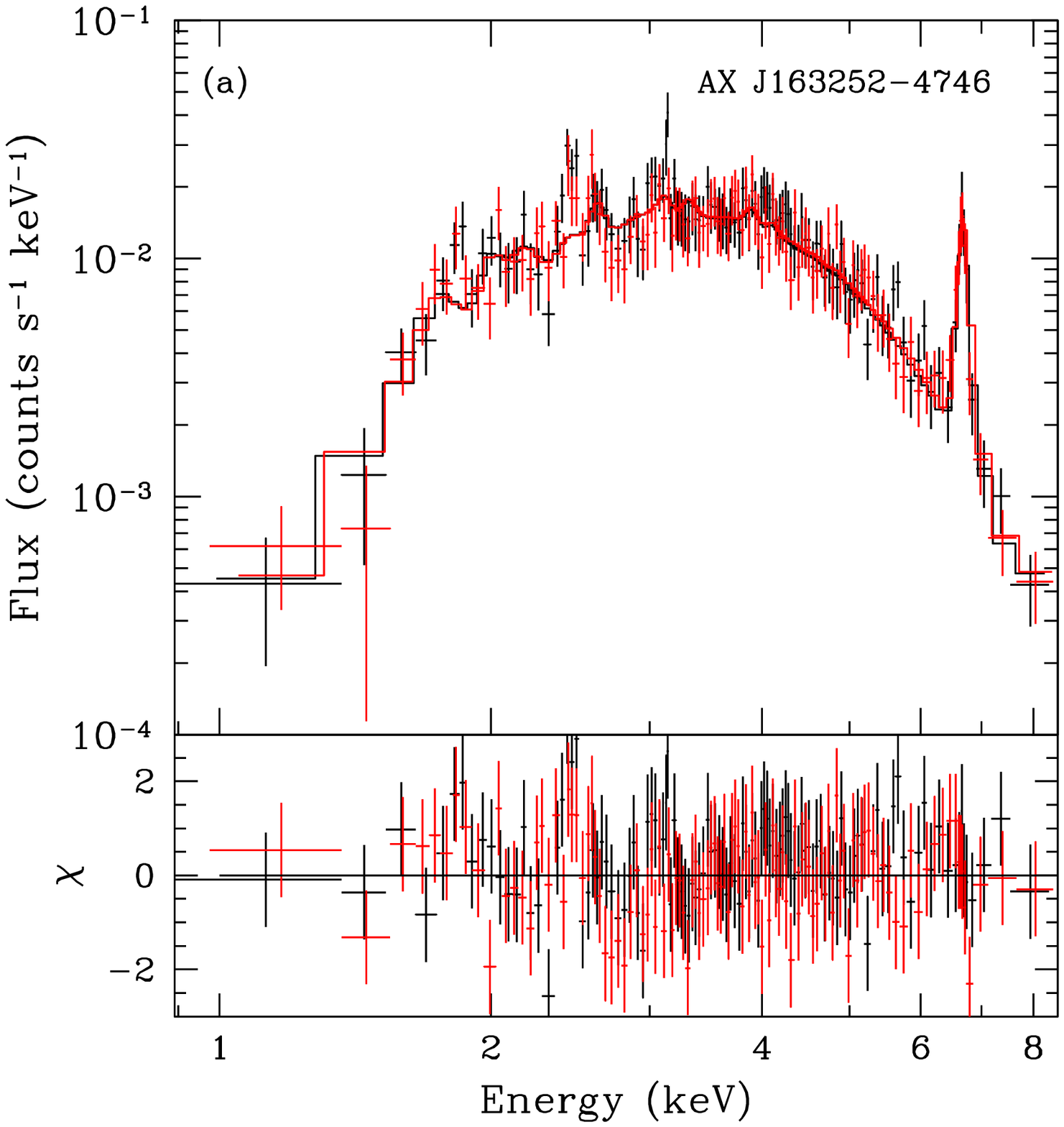}}
\subfigure{\label{f2b}\includegraphics[width=0.6\textwidth]{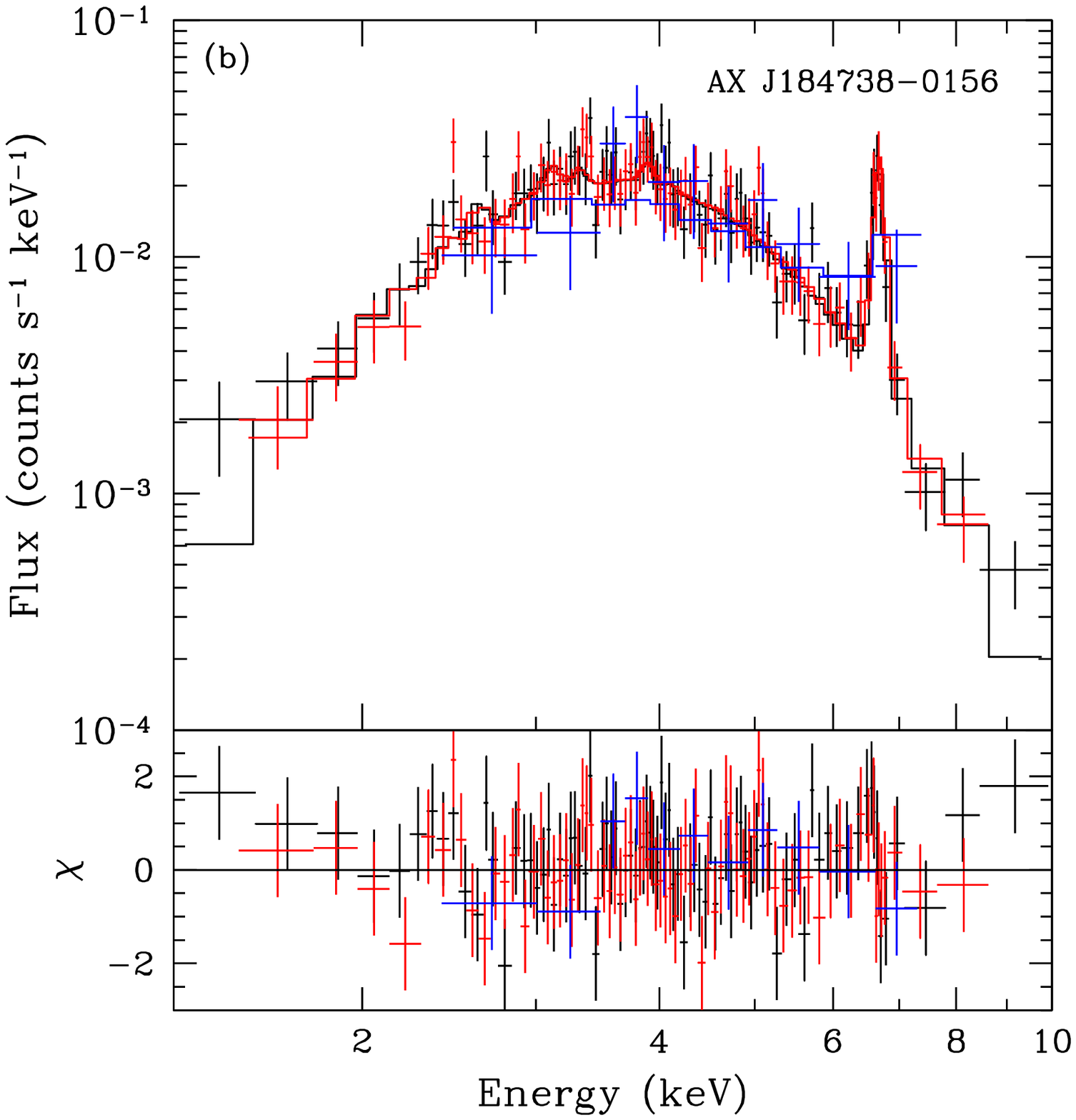}}
\caption{X-ray spectra of AX J163252--4746 and AX J184738--0156. The \xmm\ EPIC-MOS1 (black) and EPIC-MOS2 (red) spectra of these sources (data points) are overlaid with an absorbed Raymond-Smith fit (solid lines). The bottom panel in both a) and b) show the residuals of the best-fit Raymond-Smith plasma for each detector. a) \xmm\ spectra of AX J163252--4746. b) \xmm\ spectra of AX J184738--0156 overlaid with the \cxo\ (blue) spectrum of this source. The solid blue line is the \cxo\ spectral fit using the best-fit parameters from the absorbed Raymond-Smith fit to the \xmm\ data but allowing the normalization to vary. The \cxo\ residuals from this fit are also seen in the bottom panel.}
\label{fig2}
\end{center}
\end{figure}

We chose to model the \xmm\ spectra of AX J163252--4746 and AX J184738--0156, rather than their \cxo\ data, as the HRC instrument does not provide any spectral information for AX J163252--4746 and the \xmm\ observations detected many more source counts for AX J184738--0156. We fit the \xmm\ spectra of AX J163252--4746 and AX J184738--0156 with absorbed \citet{raymond77} models as both sources show a strong emission line between $6-7$ keV, indicating there is likely a thermal component to the X-ray emission. The \xmm\ spectra of both AX J163252--4746 and AX J184738--0156 are shown in Figure~\ref{fig2}.\footnote{The pipeline-produced archival \xmm\ spectra were fit using \texttt{XSPEC} \citep{dorman01}. We note that in the \xmm\ observation of AX J163252--4645, many scatter arcs from the nearby bright X-ray binary 4U~1630--47 are present. This could lead to poor background subtraction in the \xmm\ reduction pipeline. However, the other \xmm\ archival observations of AX J163252--4746, which do not have scatter arcs, show very similar spectral shapes.} Using a Gaussian profile to model the emission line in the \xmm\ spectra, the line was found to be unresolved for both AX J163252--4746 and AX J184738--0156. In each case the equivalent width (EW) is $\sim 1.3^{+0.1}_{-0.2}$ keV, at energies $6.67 \pm\ 0.01$ and $6.66 \pm\ 0.01$ keV (90\% confidence) for AX J163252--4746 and AX J184738--0156, respectively, where the \xmm\ energy resolution is FWHM $\sim$135--140 eV around 7 keV.\footnote{http://www.mssl.ucl.ac.uk/www\_xmm/ukos/onlines/uhb/XMM\_UHB/node28.html} We identify both emission lines as \ion{Fe}{25}, which has an approximate energy of 6.7 keV. The absorbed Raymond-Smith modeling of both AX J163252--4746 and AX J184738--0156 includes smaller bumps in the spectral fits, as seen in Figure~\ref{fig2}, which could indicate the presence of other spectral lines. However, the error bars in the data points are too large for the existence of other ionic species to be confirmed.

There is no evidence for short-term variability within our \cxo\ and archival \xmm\ observations between $\sim330$s and $\sim51000$s, $\sim210$s and $\sim25100$s, $\sim180$s and $\sim1450$s, and $\sim320$s and $\sim2550$s for AX J163252--4746, AX J184738--0156, AX J144701--5919 and AX J144547--5931, respectively. There is also little evidence for long-term variability as the derived fluxes from powerlaw fits to the \cxo\ and \xmm\footnote{We also included a Gaussian profile component to the powerlaw fit of the \xmm\ data in order to model the flux in the 6.7 keV \ion{Fe}{25} emission line.} observations of each source are within a factor of two from the original \asca\ flux, also derived from a powerlaw fit, published by \citet{sugizaki01}. Using the \cxo\ and \xmm\ datasets we also determined that there is no evidence for periodicity between $\sim$0.02--17000s, $\sim$0.15--8400s, $\sim$6.4-480s and $\sim$6.4-850s in AX J163252--4746, AX J184738--0156, AX J144701--5919 and AX J144547--5931, respectively, for a 99.9\% confidence.

In order to determine the validity of spectral fits applied to the \cxo\ spectra of AX J144701--5919 and AX J144547--5931 we compared the \xmm\ and \cxo\ spectra of AX J184738--0156. We used the \xmm\ best-fit absorbed Raymond-Smith values to fit the \cxo\ spectrum of AX J184738--0156, allowing the normalization parameter to vary. The resulting fit is shown in blue in Figure~\ref{f2b} demonstrating that the \cxo\ spectrum is compatible with the \xmm\ fits. An independent absorbed Raymond-Smith fit to the \cxo\ spectrum results in best-fit values that have very large uncertainties. It is therefore difficult to start from the low-statistics of the \cxo\ data and then constrain the parameters of the source's spectrum. We therefore limit our interpretation of the AX J144547--5931 and AX J144701--5919 \cxo\ spectra by using absorbed Raymond-Smith fits only to estimate the unabsorbed X-ray flux.\footnote{The \cxo\ spectra were fit using the \texttt{CIAO} spectral fitting tool \texttt{Sherpa}.} With this caveat, the Raymond-Smith spectral fit parameters for all four sources are summarized in Table~\ref{tab:2}.

\subsection{Multi-wavelength Data}

\subsubsection{Survey and Catalog Comparisons}

Comparison of the \cxo\ positions of AX J163252--4746, AX J184738--0156, AX J144701--5919 and AX J144547--5931 to optical and infrared catalogs allow us to discover any multi-wavelength counterparts to these X-ray sources. AX J144701--5919 and AX J144547--5931 are both faintly detected at optical wavelengths (R magnitude $>15.8$) in the US Naval Observatory (USNO) B catalog, version 1.0 \citep{monet03}. In both the Two Micron All Sky Survey (2MASS) \citep{skrutskie06} and the Galactic Legacy Infrared Mid-Plane Survey Extraordinaire (GLIMPSE) \citep{benjamin03} all four AGPS sources are shown to have clear counterparts at near-infrared and mid-infrared wavelengths. The corresponding 2MASS and GLIMPSE source names and magnitudes are listed in Table~\ref{tab:3}.

The GLIMPSE data also show that AX J163252--4746 is surrounded by a diffuse shell of 8$\mu$m emission (see Figure~\ref{f1a}). This emission is likely due principally to polycyclic aromatic hydrocarbons (PAHs) excited by soft ultraviolet radiation, detected within the broadband 8$\mu$m filter \citep{watson08}. Thus, such shells are signposts for hot stars with powerful stellar winds that are impacting the interstellar medium \citep{churchwell06,churchwell07}.

Examination of the 24{\micron} mosaic images from the MIPSGAL Survey \citep{carey09} shows that AX J163252--4746, AX J144701--5919 and AX J144547--5931 are also detected at this wavelength. AX J184738--0156 is situated within the H {\sc ii} region W43 \citep{lester85}. W43 appears very bright and diffuse at 24~{\micron}, preventing us from identifying a possible counterpart to AX J184738--0156 at this wavelength.

\subsubsection{Targeted Infrared Observations}

AX J184738--0156 is coincident with 2MASS~J18473666--0156334, which \citet{blum99} resolved into a dense \object[NAME W 43 STAR CLUSTER]{stellar cluster embedded within W43}. We re-observed this cluster in the $J$, $H$ and $K_{s}$ bands using the Persson's Auxiliary Nasmyth Infrared Camera \citep[PANIC:][]{martini04,osip08} on the Baade, 6.5m, Magellan telescope, located at Las Campanas Observatory, on June 25, 2007. As shown in Figure~\ref{f1b}, the X-ray source AX J184738--0156 is coincident with the brightest member of the star cluster. \citet{blum99} obtained a $K$-band spectrum of this star at a spectral resolution of $R\approx800$, classifying it as a WR star of subtype WN7; this star is now designated \object{WR 121a} \citep{vanderhucht01}.

We obtained $K$-band $2.03-2.30$ {\micron} spectra of the counterparts to AX J163252--4746, AX J144701--5919 and AX J144547--5931 on June 15 and 16, 2008 and July 20, 2009, respectively, at the 4.1m Southern Observatory for Astrophysical Research (SOAR) Telescope, located on Cerro Pachon in Chile. The Ohio State Infrared Imager/Spectrometer \citep[OSIRIS;][]{depoy93} was used to observe AX J163252--4746 and AX J144701--5919 in the high-resolution longslit mode, which provides a spectral resolution of  $R\approx3000$ in the $K$-band. Stellar spectra were acquired in a slit-scan sequence of 5 positions separated by 5\arcsec~each. We used the same instrument to observe AX J144547--5931 in the cross-dispersed mode at a resolution of $R\approx1200$, using a $10''$ slit scan with 4 repetitions. Standard reduction procedures, including telluric correction, were applied to the data.

The $K$-band spectrum of AX J163252--4746, shown in Figure~\ref{f3a}, exhibits an extremely strong emission line of He {\sc i} $\lambda$ 2.058 {\micron}, as well as strong emission features from the $\lambda$ 2.112--2.115 {\micron} complex, composed of He {\sc i}, N {\sc iii}, C {\sc iii} and O {\sc iii}, and Br{$\gamma$}/He {\sc i} $\lambda$ 2.166 {\micron}.  The lower panel of Figure~\ref{f3a} is a magnified view of this star's spectrum showing the presence of weaker emission features from N {\sc iii} $\lambda$2.247 {\micron}, He {\sc ii} $\lambda$2.189 {\micron} and possibly N {\sc v} at $\lambda$2.100 {\micron}.

The $K$-band OSIRIS spectra of AX J144701--5919 and AX J144547--5931 are shown in Figure~\ref{f3b}. While both AX~J144701--5919 and AX J144547--593 show clear emission lines, the spectrum of the latter is poorer, as high clouds and vibrations due to strong wind led to a high background and significant flux variation between the exposures. The $K$-band spectrum of AX J184738--0156 can be found in Figure~5 of \citet{blum99}. In all three cases the $\lambda$ 2.112--2.115 complex {\micron} and Br{$\gamma$}/He {\sc i} $\lambda$ 2.166 {\micron} are seen in emission. AX J144701$-$5919 and AX J184738-0156 also have a strong He {\sc ii} emission line at $\lambda$ 2.189 {\micron}. While AX J184738-0156 shows He {\sc i} $\lambda$ 2.058 {\micron} in emission, for AX J144701--5919 this line has a P-Cygni profile, with a weak emission, but strong absorption component. AX J144547--0593 displays the C~{\sc iv} $\lambda$ 2.070-2.084 {\micron} emission line complex. The atomic transitions and center wavelengths of these emission lines can be found in \citet{morris96}.  

\subsubsection{Targeted Optical Observations}

A low signal-to-noise optical spectrum of AX J144547--5931 was obtained with the Low Dispersion Survey Spectrograph (LDSS-3) \citep{osip08} on the 6.5m Clay Magellan telescope, on June 23, 2008. This source was observed using the VPH\_ALL (400 lines/mm) grism, with a center slit, resulting in a wavelength range of $3000-11000$ {\AA}. This spectrum shows H$\alpha$ in emission, with C {\sc iii} $\lambda$ 5696 and C {\sc iii} $\lambda$ 6721, 6727, 6731 {\AA} emission lines. Figure~\ref{f3c} displays the part of the LDSS-3 spectrum with these emission lines.

\subsubsection{Radio Observations}

AX J163252--4746, AX J144701--5919 and AX J144547--5931 are coincident with radio emission in the first and second epoch Molonglo Galactic Plane Surveys \citep[MGPS1 and MGPS2 respectively;][]{green99,murphy07}. The MGPS data sets are made up of mosaic images, taken at 843 MHz with the Molonglo Synthesis Telescope (MOST). 

AX J144701--5919 appears to be associated with an unresolved radio source. Molonglo observed this source in 1994 and 1998, with a measured flux density of $15 \pm\ 5$ and $10 \pm\ 5$ mJy respectively. Within these large errors there is no evidence of flux variability between these epochs. AX J163252--4746 and AX J144547--5931 fall within regions of extended radio emission. AX J184738--0156 is coincident with a bright knot within diffuse radio emission from the surrounding H {\sc ii} region W43, as seen at 1.4 GHz in the Multi-Array Galactic Plane Imaging Survey \citep[MAGPIS:][]{helfand06}.

We used the Australia Telescope Compact Array (ATCA) to follow-up the counterpart to AX J144701--5919 and resolve out the diffuse radio emission surrounding AX J163252--4746, observing each source for one hour at each of 1.4, 2.4, 4.8 and 8.6 GHz with a 6km baseline configuration on January 21, 2008 (AX J144701--5818) and April 11, 2008 (AX J163252--4746).\footnote{During the observation of AX J163252--4746 the antenna providing the 6km baseline was not operational at 1.4 and 2.4 GHz. This resulted in 3km being the longest baseline during the observations of this source at these frequencies.} The radio counterpart to AX J144701--5919 was detected as an unresolved point source in all 4 ATCA frequency bands, with a flux that decreased with frequency according to $S^{total}_{\nu} \approx \nu^{\alpha}$ with $\alpha = -0.54 \pm 0.09$. AX J163252--4746 was only detected at 4.8 and 8.6 GHz resulting in a spectral index of $\alpha = +0.19 \pm 0.02$. The observed ATCA radio fluxes and upper limits for AX J163252--4746 and AX J144701--5919 are listed in Table~\ref{tab:3}. Further high-resolution radio follow-up is required to determine whether AX J184738--0156 and AX J144547--5931 have compact radio counterparts.

\section{Discussion}

\subsection{Classification of Stellar Counterparts}

The infrared colors of AX J163252--4746, AX J184738--0156, AX J144701--5919 and AX J144547--5931 are reasonably consistent with the \citet{hadfield07} and \citet{mauerhan09} selection criteria for WR stars, i.e., GLIMPSE colors $[3.6]-[8.0]>0.5$ and $[3.6]-[4.5]>0.1$ and a 2MASS+GLIMPSE color of $K - [0.8] > 0.7$. This color selection criteria takes advantage of the free-free excess emission that is generated within the strong, dense, ionized winds of WRs. The colors of our four target, thus, indicate the presence of such winds. The PAH ring surrounding AX J163252--4746 (Figure~\ref{f1a}) is also consistent with a strong wind being generated by the central source.

The $K$-band spectrum of AX J184738--0156 obtained by \citet{blum99} and our OSIRIS $K$-band spectra of AX J163252--4746, AX J144701--5919 and AX J144547--5931 all show strong emission lines from the $\lambda$ 2.112--2.115 {\micron} complex and Br{$\gamma$}/He {\sc i} $\lambda$ 2.166 {\micron}. The presence of such emission lines are indicative of these stars being massive ($>20 M_{\odot}$), from either the late-type WR stars in the nitrogen sequence (WN) that are hydrogen-rich or their O-type supergiant progenitors, particularly of the OIf$^{+}$ variety \citep{martins08}. WR is the name traditionally given to hydrogen-poor, emission line stars, however, there is a subset of the nitrogen sequence WR stars that have hydrogen in their spectra. \citet{smith08} proposed a new designation for this subset, WNH, as they are different from classical WR stars in that they are still undergoing hydrogen-core burning and are therefore at an earlier stage in their evolution. WNH stars are some of the most massive in our Galaxy, with a mass distribution that peaks around $\sim50$ $M_{\odot}$, but which can reach as high as $120$ $M_{\odot}$ \citep{smith08}, making them very short-lived and rare. They can have luminosities as high as $2 \times 10^{6}$ $L_{\odot}$ and large mass-loss rates of $\dot{M} > 10^{-5}$ $M_{\odot}$yr$^{-1}$ due to fast extended winds \citep{martins08}. OIf$^{+}$ stars are also extremely massive with very similar properties \citep{martins08}. Both WNH and OIf$^{+}$ stars are known X-ray emitters \citep[e.g.][]{mauerhan10}.

\begin{turnpage}
\begin{figure}
\begin{center}
\subfigure{\label{f3a}\includegraphics[width=0.47\textwidth]{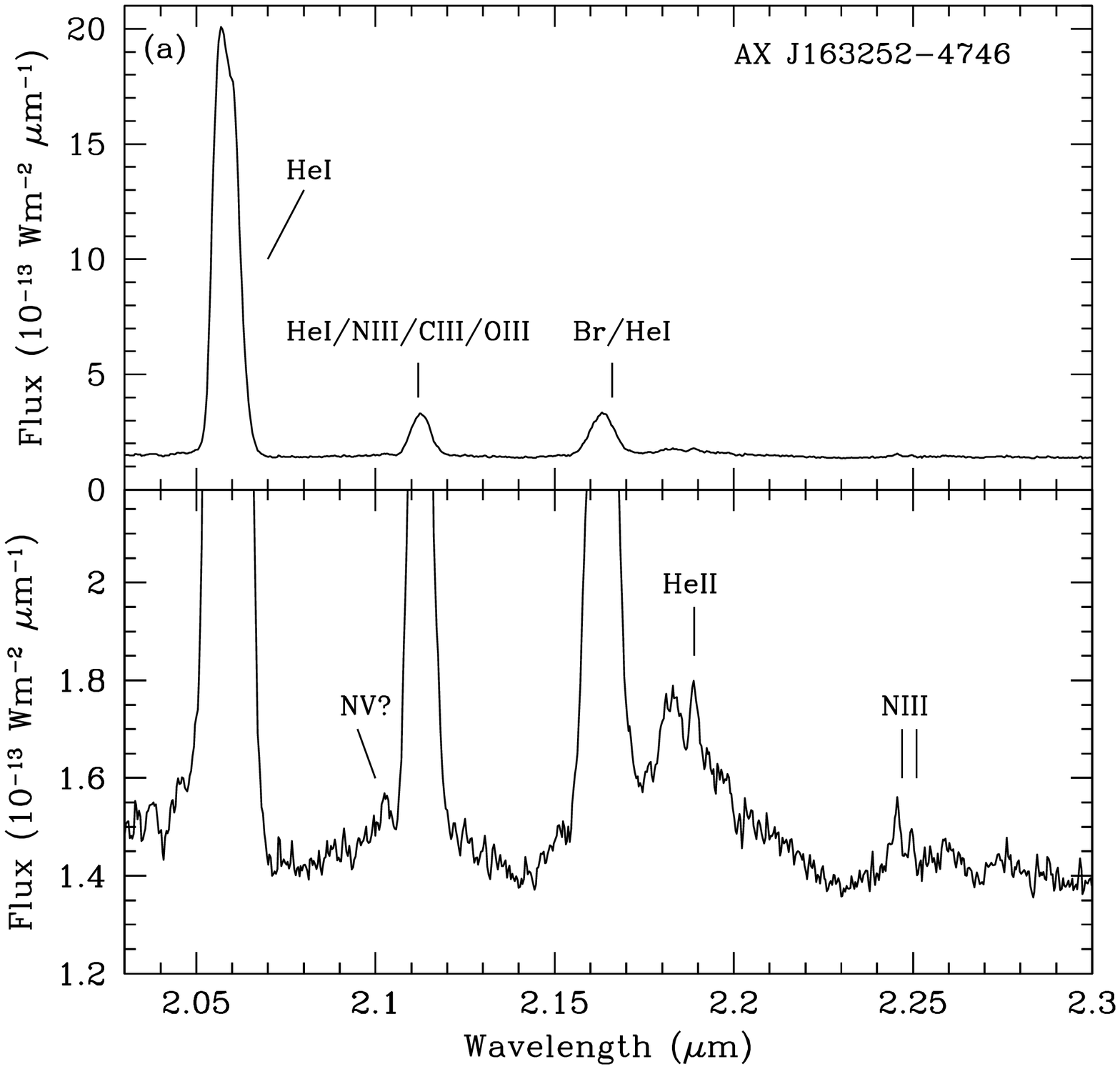}}
\subfigure{\label{f3b}\includegraphics[width=0.47\textwidth]{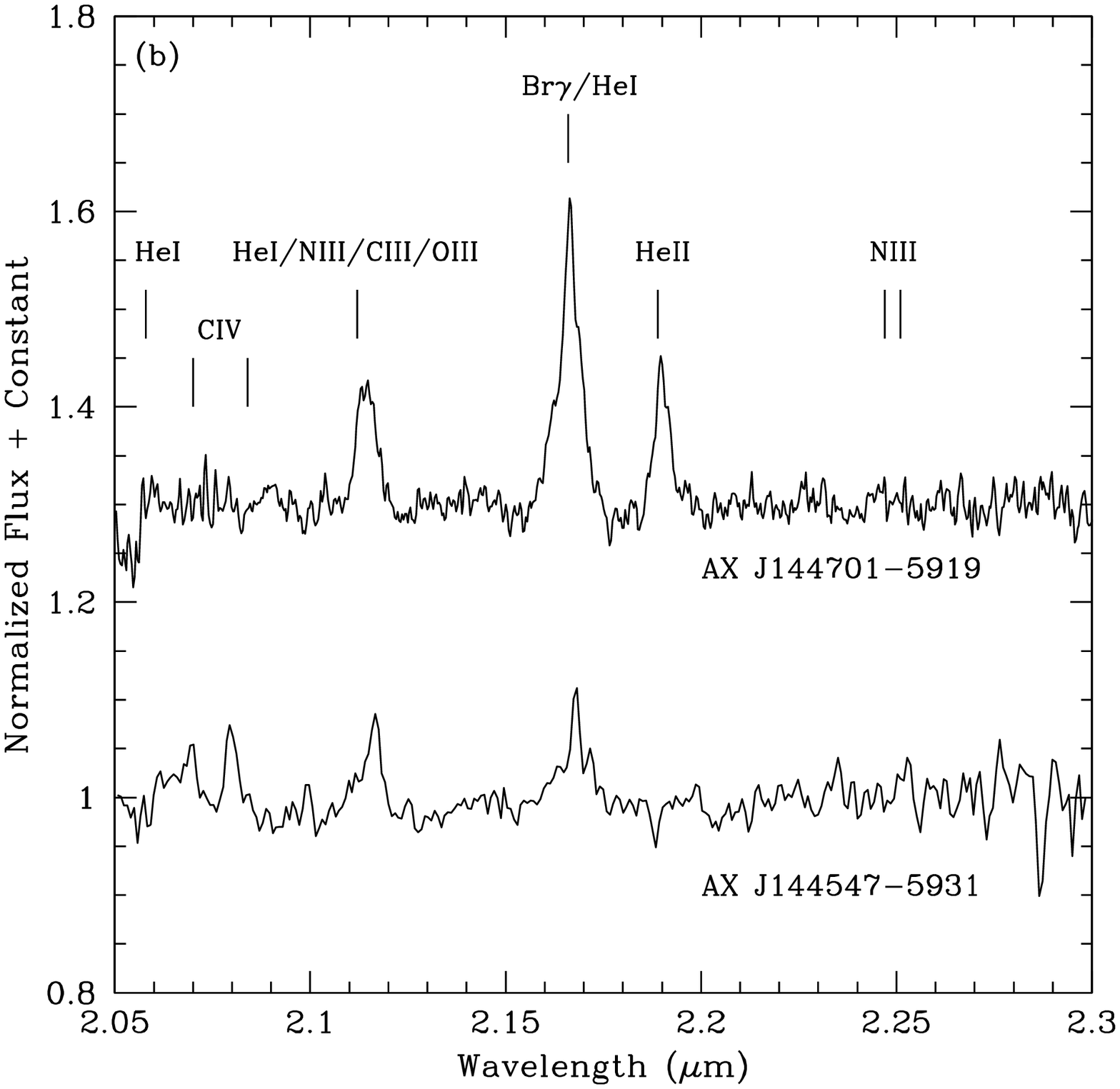}}
\subfigure{\label{f3c}\includegraphics[width=0.47\textwidth]{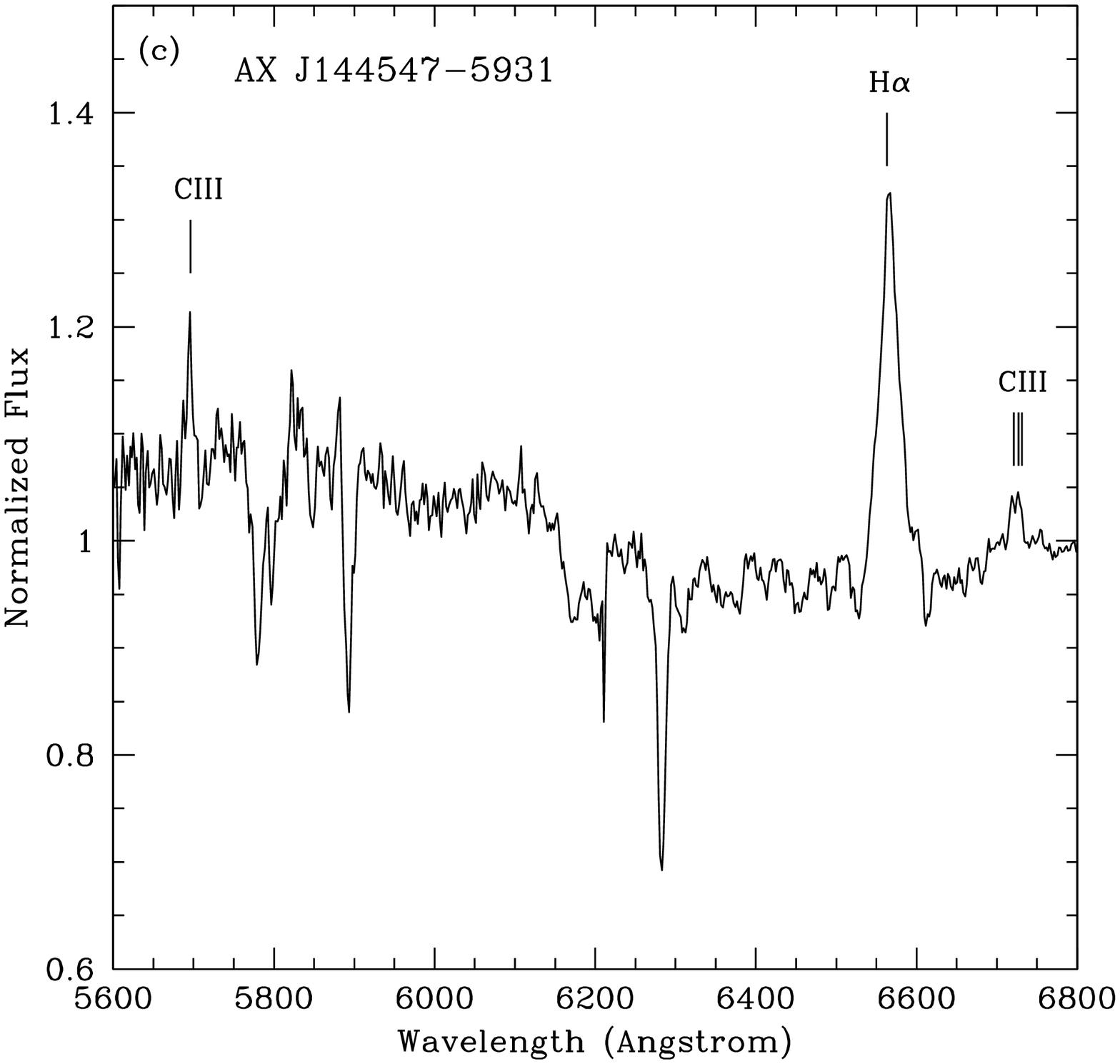}}
\caption{Infrared and optical spectra of AX J163252--4746, AX J144701--5919 and AX J144547--5931. a) OSIRIS $K$-band spectrum of AX J163252--4746. The upper panel shows the full amplitude of the flux whereas the lower panel is a magnified view of the spectrum, indicating weaker emission lines. b) OSIRIS $K$-band spectra of AX J144701--5919 and AX J144547--5932. c) LDSS3 optical spectrum of AX J144547--5931.}
\label{fig3}
\end{center}
\end{figure}
\end{turnpage}

WNH and OIf$^{+}$ stars also produce other spectral line features allowing us to distinguish between the different subtypes. The $K$-band spectrum of AX J163252--4746 (Figure~\ref{f3a}) is remarkable, in that it exhibits an ultra-strong He {\sc i} $\lambda$ 2.058 {\micron} emission line. The total equivalent width is $\approx$1040 {\AA}, with a FWHM of $\approx$74 {\AA} (1080 km s$^{-1}$). The spectrum also exhibits strong emission features from the $\lambda$2.112--2.115 {\micron} complex ($\textrm{EW}\approx80$ \AA) and Br{$\gamma$}/He {\sc i} $\lambda$2.166 {\micron} ($\textrm{EW}\approx90$ \AA). The relative strengths of the emission lines, especially the very strong He {\sc i} emission, are similar to those of the Ofpe/WN9 stars \object{WR 122} and \object{WR 85a} \citep{morris96,figer97}, although the individual line strengths are significantly stronger for AX J163252--4746 than for these other sources. Ofpe/WN9 stars, also known as ``slash stars,'' have been described as both WNHs or luminous blue variables \citep{smith08}. We assign AX J163252--4746 the spectral type Ofpe/WN9.

The zoomed view of AX J163252--4746, seen in the lower panel of Figure~\ref{f3a}, show fainter emission features that could indicate the presence of a massive companion. There is a spectral hint of broad He {\sc i} $\lambda$ 2.112 {\micron} emission under the narrower line of the brighter star and a broad spectral feature from He {\sc ii} $\lambda$ 2.189 {\micron}. Though He {\sc ii} $\lambda$ 2.189 {\micron} is a known emission feature of Ofpe/WN9 stars, the broadness of this line and the He {\sc i} feature are both reminiscent of an early-type WN star (WNE) of the WN4--6 variety \citep{figer97}. If these features are real, their weakness may be due to the fact that WNE stars are up to two magnitudes fainter than Ofpe/WN9 stars, intrinsically \citep{crowther06}. The questionable weak feature near $\lambda$ 2.10 {\micron}, speculated to be N {\sc v} emission, could also indicate the presence of an underlying WNE spectrum since this line requires the associated high temperatures of a WNE star. The weak N {\sc iii} $\lambda$ 2.247 {\micron} doublet could be from either star, though is more ordinarily seen in late-type WNs, which include the WN7--9 subtypes \citep{figer97,martins08}. Finally, as a separate speculation, we note that the shape of the peak in the He {\sc i} $\lambda$ 2.058 {\micron} line, seen in the top panel of Figure~\ref{f3a}, appears to have a double peaked profile, as if it were a blend of two velocity-shifted components having comparable line strengths. However, none of the other strong lines exhibit this profile. Given the current data this is all very speculative. We defer any definitive conclusions able the weak spectral features or the potential contribution of a companion spectrum until further observations are conducted. Higher resolution spectroscopic measurements over several epochs will be required to determine if the weak, underlying, broad features are real, and to identify this source as a spectroscopic binary. 

As noted in $\S 2.2.2$, AX J184738--0156 is coincident with WR 121a, which was identified by \citet{blum99} as a star of subtype WN7. The $K$-band spectrum of this star, seen in Figure~5 of \citet{blum99}, is similar to WN7--8h stars, which are WNH subtypes \citep[e.g. see][]{martins08,mauerhan10}. X-ray emission has not been connected with WR 121a prior to this paper. However, there is evidence to suggest that WR 121a and its surrounding star-formation region, W43, may be associated with the extended (intrinsic rms size of $0.32^{\circ} \pm 0.02^{\circ}$) very-high-energy (E $> 100$GeV) $\gamma$-ray source \object[HESS J1848-018]{HESS~J1848--018} \citep{chaves08} and its possible counterpart, \object[0FGL J1848.6-0138]{0FGL J1848.6--0138}, detected with the Large Area Telescope (LAT) onboard the \textit{Fermi Gamma-ray Space Telescope} \citep{tam10}. W43 is also host to the bright H{\sc ii} region G30.8--0.2 and a giant molecular cloud, which partially overlaps with HESS J1848--018, their centroids separated by $\sim0.3^{\circ}$. WR 121a lies within the region of this HESS source's extended emission, $\sim0.2^{\circ}$ from its centroid \citep{chaves08}. A similar offset exists between HESS J1023--575 and the young open cluster, Westerlund 2, host to WR 20a \citep{aharonian07} yet the extended nature of this HESS source, and therefore HESS J1848--018, argues against a single star origin for these VHE $\gamma$-ray sources \citep{aharonian07,chaves08}. 

For AX J144701--5919 the Br{$\gamma$}/He {\sc i} emission line in the $K$-band spectrum (Figure~\ref{f3b}) is stronger than the $\lambda$2.112--2.115 {\micron} emission line complex indicating a spectral type later than WN6. As the He {\sc ii} $\lambda$ 2.189 {\micron} feature is also seen in emission we expect a spectral type earlier than WN8-9 as this feature would otherwise appear in absorption or would have a P-Cygni profile \citep[e.g., see][]{martins08}. The ratio of equivalent widths of the detected lines can be used to quantitatively determine the WN subtype. We measure EW ($\lambda$2.189 {\micron})/EW($\lambda$2.112 {\micron}) = 0.4 and EW ($\lambda$2.189 {\micron})/EW($\lambda$2.166 {\micron}) = 0.3, which are most consistent with WN7--8 subtypes \citep{figer97,martins08}. We therefore classify the counterpart to AX J144701--5919 as a WNH star of subtype WN7--8h.

The $K$-band spectrum of AX J144547--5931 (Figure~\ref{f3b}) shows He {\sc ii} $\lambda$ 2.189 {\micron} in absorption, which is indicative of either a WN8--9 or an OIf$^{+}$ star \citep{martins08}. As the $\lambda$ 2.112--2.115 {\micron} complex and the Br{$\gamma$}/He {\sc i} $\lambda$ 2.166 {\micron} emission line are of similar strength it is more likely an OIf$^{+}$ subtype \citep{martins08}. The LDSS-3 optical spectrum of AX J144547--5931 (Figure~\ref{f3c}) is missing the He {\sc i} $\lambda$ 6678 {\micron} emission line seen very strongly in optical spectra of WN8--9 stars \citep[e.g., see][]{corradi10}. It does have a very broad H$\alpha$ line, as well as C {\sc iii} $\lambda$ 5696 {\AA} and C {\sc iii} $\lambda$ 6721, 6727, 6731 {\AA} emission, which are known O-type star spectral lines \citep{walborn01}. These spectral features are characteristic of OIf$^{+}$ stars identified by \citet{becker06,becker09}. Further spectroscopic follow-up is required to constrain the OIf$^+$ subtype of AX J144547--5931. This could be achieved by the detection of Si {\sc iv} emission at $\lambda$ 4089--4116 {\AA}, which is, historically, the defining characteristic of high-luminosity OIf$^{+}$ stars \citep{walborn71}.

\subsection{Distances and X-ray Luminosities}

Stars with the spectral classification of AX J163252--4746, Ofpe/WN9, have a narrow range in absolute $K$-band magnitude ($M_{K}$), which can be combined with an estimate of the extinction along the line of sight to determine the approximate distance to the source. This method is outlined and tabulated by \citet{hadfield07} and \citet{mauerhan09}, and adopts intrinsic colors from \citet{crowther06}. Assuming the colors and $M_{K}$ value for a WN10--11 star (aka Ofpe/WN9) from \citet{crowther06} we derive an approximate distance of 4.9 kpc. Using red clump stars to measure the reddening as a function of distance along the line of site \citep{lopez02} we can also calculate a lower limit of $\gtrsim4.5$ kpc on the distance to AX J163252--4746, which is consistent with our result. We therefore adopt a distance of $\sim4.9$ kpc to AX J163252--4746 yielding an unabsorbed X-ray luminosity of $L_{x} \approx 3.4 \times 10^{34}$ erg s$^{-1}$ in the $0.3-8.0$ keV energy range.

The kinematic distance to W43, in which AX J184738--0156 is embedded, is 5.6 kpc \citep{smith78}.\footnote{Revised for a modern distance of $\sim8$ kpc to the Galactic center \citep{reid09}.} This is also consistent with the red clump lower limit of $\gtrsim3.5$ kpc. For a distance of 5.6 kpc to AX J184738--0156 we calculate an unabsorbed luminosity of $L_{x} \approx 3.4 \times 10^{34}$ erg s$^{-1}$ in the $0.3-8.0$ keV energy range.

AX J144701--5919 and AX J144547--5931 are separated on the sky by only $\sim16'$ and are coincident with a star formation complex. There are close surrounding H {\sc ii} regions, \object[GAL 316.81-00.04]{G316.808--0.037}, \object[GAL 317.04+00.30]{G317.037+0.300} and \object[GAL 317.29+00.27]{G317.291+0.268}, for which \citet{caswell87} used radio recombination line velocities to calculate a near and a far kinematic distance to each.  As \citet{shaver81} present radio and infrared observations indicating that G316.808--0.037 is located at the closer kinematic distance of $\sim2.3$ kpc we assume the nearer distance for the other two regions also, so that G317.037+0.300 and G317.291+0.268 lie at $\sim2.8$ and $\sim2.9$ kpc, respectively. Using red clump stars to measure the reddening as a function of distance along the line of site \citep{lopez02} we can calculate a lower limit of $\gtrsim3.0$ kpc to these two X-ray sources, which is reasonably consistent with the kinematic distances.

Stars with the spectral classification of AX J144701--5919, WN7--8h, also have a narrow range in absolute $K$-band magnitude. By averaging together the $M_{k}$ values of two WN7--8h stars in the Arches cluster \citep{martins08} and using the \citet{crowther06} colors for WN7--9 stars to estimate the extinction along the line of sight, we further refine our distance estimate to $\sim2.6$ kpc for AX J144701--5919, resulting in an unabsorbed luminosity of $L_{x} \approx 3.2 \times 10^{33}$ erg s$^{-1}$ in the $0.3-8.0$ keV energy range. This distance is fairly consistent with the red-clump value and estimates via comparison with neighboring H {\sc ii} regions. 

As we do not have a strong enough constraint on the OIf$^{+}$  luminosity class we cannot apply the same method to AX J144547--5931. We therefore derive a range of possible unabsorbed X-ray luminosities based on both the kinematic and red clump distances, $2.3-3.0$ kpc. In this case $L_{x} \approx 1.9 - 3.2 \times 10^{33}$ erg s$^{-1}$ in the $0.3-8.0$ keV energy range. 

In each case the unabsorbed X-ray luminosity is derived from the absorbed Raymond-Smith spectral fits to the X-ray spectra. These luminosities can be found in Table~\ref{tab:2} along with the distances used for the calculations. All these distances have been revised for a modern distance of 8 kpc to the Galactic center \citep{reid09}.

\subsection{AX J163252--4746 and AX J184738--0156}

AX J163252--4746 and AX J184738--0156 are the two most luminous sources out of the four investigated in this paper. We discuss these sources together as they are both associated with WNH stars and have similar X-ray luminosities and characteristics as well as comparable observational information. There are three possibilities as to how these WNH stars can produce the X-ray emission we observe. These include X-rays that are intrinsic to the massive star, X-rays produced through gravitation accretion in a HMXB or X-rays from shock-heated gas in a CWB. We now discuss each possibility in detail.

Many WN stars have been shown to have thermal X-ray emission created through instability-driven wind-shocks with typical temperatures around $kT \approx 0.6$ keV \citep{oskinova05}. However, the \xmm\ spectra of AX J163252--4746 and AX J184738--0156 (Figure~\ref{fig2}) show that few X-ray photons were detected with energies $<1$ keV, indicating high levels of absorption preventing the detection of such a stellar wind component. Raymond-Smith fits to the same spectra result in temperatures of $kT =3.2$ and 3.1 keV for AX J163252--4746 and AX J184738--0156, respectively (see Table~\ref{tab:2}), so the majority of the observed X-ray emission from these two sources likely arises from another mechanism capable of generating a much hotter plasma. 

In an isolated massive star thermal X-rays can also be generated through the magnetically channeled wind-shock mechanism. \citet{gagne05} have successfully demonstrated that this model can explain the high temperature ($kT \sim3$ keV) X-ray plasma and X-ray luminosity ($L_{x} \approx 10^{33}$ erg s$^{-1}$; $0.5-10$ keV) associated with the O5.5 V star \object[HD 37022]{$\theta^{1}$ Orionis C}. However, there are problems applying this same model to WR stars as they have far stronger wind momenta than O stars. \citet{mauerhan10} showed that the WR stars and O supergiants in their sample, which have similar X-ray properties and luminosities to the four sources discussed in this paper, would require $\approx 5$ kG magnetic fields to confine the winds. Thus far no such fields have been detected around WR stars. We thus regard this scenario as very unlikely.

The generation of non-thermal X-rays in a single massive star has also been considered \citep[e.g.][]{chen91}, however, recent studies of O stars have shown that in most cases, at energies $<10$ keV, any non-thermal X-rays are likely to make an insignificant contribution to the overall stellar X-ray spectrum \citep{pittard10}. We assume that this is also the case for WNH stars as they are only slightly more evolved than massive O stars \citep{smith08}. 

Given the above arguments it is unlikely that the majority of X-ray emission from either AX J163252--4746 or AX J184738--0156 is generated through the instability-driven or magnetically channeled wind-shock mechanisms, or through inverse-Compton scattering. The X-rays are therefore not intrinsic to the WNH counterparts, so further scenarios need to be considered.

The radio spectrum of a single WR star with an optically thick stellar wind is thermal and has a spectral index of $\alpha \approx +0.7 \pm\ 0.1$ \citep{dougherty00}. If the wind is optically thin and has a radio spectral index $<+0.6$ this implies that there is another component to the emission. The ATCA observation of AX J163252--4746 shows its radio spectrum to be approximately flat ($\alpha \approx +0.2$). This could indicate that there is a significant non-thermal component to the spectrum where this and the thermal component of the source are inferred to have similar luminosities. Such spectra are referred to as ``composite'' \citep{dougherty00}. Likely origins of non-thermal radio emission associated with massive stars are either synchrotron emission from ejected material \citep[][and references therein]{fender01} or from short-lived jets \citep[e.g.,][]{pestalozzi09} in HMXBs or relativistic electrons accelerated in the wind collision region (WCR) of a CWB \citep{eichler93,dougherty03,pittardIC06}. A flat spectral index may also occur if the optically thin thermal radio emission from the WCR of a CWB is comparable to that of the thermal emission from the unshocked winds in the system \citep{pittardIC06}. The resulting spectrum would resemble that of a composite spectrum.

Of the two main classes of HMXBs, AX J163252--4746 and AX J184738--0156 are more likely to be a form of SGXB given their counterpart classification as WNH stars with infrared excess implying strong stellar winds. Most SGXBs are persistent X-ray sources with X-ray luminosities around $L_{x} \sim 10^{35} - 10^{36}$ erg s$^{-1}$ \citep{sguera06}. Currently we only know of four HMXBs with WR companions, three of which are SGXB BH candidates and only one, \object[Cygnus X-3]{Cygnus X--3}, is found in the Milky Way \citep{szostek08}. The other, \object[OAO 1657-415]{OAO 1657--415}, which is also the AGPS source \object[AX J1700.7-4139]{AX J170047--4139} \citep{sugizaki01}, is an X-ray pulsar with an Ofpe/WN9 companion \citep{mason09}. This source has an X-ray luminosity of $L_{x} \sim 10^{36} - 10^{37}$ erg~s$^{-1}$ and appears to be transitioning from a wind-fed SGXB into a BeX disk-fed system \citep{mason09}. ``Classical'' SGXBs and systems such as OAO 1657--415 are therefore too luminous to explain the X-ray emission seen in AX J163252--4746 and AX J184738--0156.

However, the unabsorbed X-ray luminosities of AX J163252--4746 and AX J184738--0156 are far more consistent with the quiescent luminosities of a subclass of SGXBs that have recently been discovered with the \textit{Integral} satellite, called supergiant fast X-ray transients \citep[SFXTs:][]{sguera06}. These systems are neutron star HMXBs that produce extremely bright, rapid X-ray bursts with luminosities reaching $L_{x} \sim 10^{36}$ erg s$^{-1}$ \citep{chaty10} and have quiescent luminosities at or below $L_{x} \sim 10^{32} - 10^{33}$ erg s$^{-1}$ \citep{sguera06}. Currently very few sources with X-ray bursts have been confirmed as SFXTs, most of which contain X-ray pulsars \citep{negueruela06,sguera06}. The X-ray luminosities of AX J163252--4746 and AX J184738--0156, listed in Table~\ref{tab:2}, are at least an order of magnitude brighter than the quiescent luminosities of SFXTs and the X-ray flux observed for each source during the \asca, \xmm\ and \cxo\ observations shows no evidence for long term variability. A thorough investigation of the \xmm\ and \cxo\ observations also show no evidence of short term variability, flaring or pulsation. SFXTs also have hard, nonthermal X-ray spectra that are well described by a power law with a photon index of $\Gamma \sim 1$ \citep{lutovinov05}. Power-law fits to the \xmm\ spectra of AX J163252--4746 and AX J184738--0156 are much steeper than this with photon indices of $\Gamma \sim 3$. 

 As the \xmm\ spectra of AX J163252$-$4746 and AX J184738$-$0156 both have a strong Fe {\sc xxv} emission line at $\sim6.7$ keV, indicating the presence of very high temperature plasma, these spectra are better described by a thin thermal plasma model, not typically seen in HMXBs. While low iron ionization state emission lines (Fe {\sc i} - Fe {\sc xii}) have been seen in many HMXB X-ray spectra \citep[e.g.,][]{tomsick09}, created by cool accretion disks irradiated by a high-energy source \citep{caballero09}, it is hard to explain the presence of the $\sim6.7$ keV Fe {\sc xxv} in the X-ray spectra of these two sources.  This line can only be created in an environment capable of producing a very highly ionized plasma \citep{reig99}. Given this fact, combined with the other X-ray properties of AX J163252--4746 and AX J184738--0156, a HMXB interpretation does not seem viable.

We now consider a CWB interpretation for these two sources. The X-ray luminosities from WNH CWBs are usually between $L_{x} \sim 10^{32} - 10^{34}$ erg~s$^{-1}$ \citep{mauerhan10}. Our estimates of the unabsorbed luminosities of AX J163252--4746 and AX J184738--0156 are consistent with this range. The very hot thermal plasma ($kT > 3$ keV) seen in the Raymond-Smith fits to the \xmm\ spectra of AX J163252--4746 and AX J184738--0156 indicates binarity, where the hard X-ray emission could be generated by colliding supersonic winds in a CWB \citep{prilutskii76,cherepashchuk76,usov92}. 

The interface between the winds of two massive stars is capable of generating the hot thermal plasma necessary to produce the 6.7 keV Fe {\sc xxv} emission line seen in the \xmm\ spectra of AX J163252--4746 and AX J184738--0156. \citet{raassen03} argue that for \object{WR 25}, another WNH star classified as WN6h, the 6.7~keV Fe {\sc xxv} emission line seen in its X-ray spectrum can only be created in a wide binary, made up of this star and a massive WR or O star companion, with their winds thus colliding at speeds close to their terminal velocities of over $1000$ kms$^{-1}$ \citep{vanderhucht01}. It should be noted here that this might not be true for all WR CWBs \citep[see][]{zhekov10}. As both AX J163252--4746 and AX J184738--0156 show a lack of variability and periodicity in X-rays, have thermal X-ray plasma temperatures and luminosities similar to known WNH CWBs, and also display the 6.7~keV Fe {\sc xxv} emission line in their X-ray spectra, we classify these two sources as colliding-wind binaries.

The same relationship that we see between the X-ray and bolometric luminosities of single O stars, attributed to instability-driven wind-shocks, has been shown to hold for WN$+$O and O$+$O binaries, where $\log(L_{x}/L_{bol}) \approx\ -7$ \citep{oskinova05,mauerhan10}. This suggests that a major fraction of the X-rays from CWBs are soft ($<2.5$ keV), emitted by the individual stellar winds rather than in the WCR \citep{oskinova05,antokhin08}. That being said, there is no relationship known to exist for putatively single WN stars \citep{oskinova05,mauerhan10} so it is still perplexing as to why there are not more luminous massive binaries where large amounts of hard X-rays are generated in the WCR \citep{antokhin08}. There are therefore few cases where CWBs are overluminous, with $\log(L_{x}/L_{bol}) > -7$. We calculated a range of $\log(L_{x}/L_{bol})$ for AX J163252--4746 and AX J184738--0156 by averaging the $L_{bol}$ values of the Ofpe/WN9 stars listed in \citet{oskinova05} and \citet{mauerhan10} for AX J163252--4746 and the WN7--8h stars in \citet{mauerhan10} for AX J184738--0156. For both sources we find $\log(L_{x}/L_{bol}) \gtrsim -5.4$, as listed in Table~\ref{tab:2}. There are very few examples of WN CWBs where $\log(L_{x}/L_{bol}) > -6$ and these include \object[CXOGC J174550.2-284911]{CXOGC J174550.2--284911}, a WN8--9h CWB with $\log(L_{x}/L_{bol}) \sim -5.8$ \citep{mauerhan10}, and WR 25 (WN6ha+04f) with $\log(L_{x}/L_{bol}) \sim -5.6$ \citep{raassen03}. AX J163252--4746 and AX J184738--0156 therefore appear to be extremely X-ray luminous CWBs.

\subsection{AX J144701--5919 and AX J144547--5931}

AX J144701--5919 and AX J144547--5931 are much fainter and somewhat softer than AX J163252--4746 and AX J184738--0156 but all four sources have similar massive counterparts.  We now address the isolated massive star, HMXB and CWB scenarios for AX J144701--5919 and AX J144547--5931.

Due to the low number of counts in their \cxo\ observations we cannot obtain a reliable spectral fit or identify any X-ray emission lines in the X-ray spectra of AX J144701--5919 and AX J144547--5931. Instead we explore the hardness of their X-ray spectra by seeing how the 50\% and 75\% photon energy ($E_{50}$ and $E_{75}$, respectively), the energy below which 50\% and 75\% of the photons energies are found, of these two sources compare to the same values for AX J163252--4746 and AX J184738--0156. The $E_{50}$ value indicates whether a given source is hard ($E_{50} > 2$ keV) or soft ($E_{50} < 2$ keV) and $E_{75}$ keV grades the level of hardness within the hard category. In all four cases $E_{50}$ is between 2 and 3 keV. The $E_{75}$ value of AX J163252--4746, AX J184738--0156, AX J144701--5919 and AX J144547--5931 are  $\sim4.4$, $\sim4.5$, $\sim2.9$ and $\sim2.9$ keV, respectively. Through this crude comparison it appears that the overall spectra of AX J144701--5919 and AX J144547--5931 may be reasonably soft, consistent with the X-rays primarily being generated through instability-driven wind shocks, intrinsic to the individual stars.

The X-ray luminosities of AX J144701--5919 and AX J144547--5931, listed in Table~\ref{tab:2}, are slightly brighter than many putatively single WN and O stars \citep[e.g.][]{oskinova05,antokhin08} but are consistent with the X-ray luminosities of CWBs and quiescent SFXTs. However, using the same method as described in Section $\S 3.3$, we calculated a range of $\log(L_{x}/L_{bol})$ for both sources, listed in Table~\ref{tab:2}. In each case $\log(L_{x}/L_{bol}) > -6.5$ making these sources overluminous compared to the $\log(L_{x}/L_{bol}) \approx -7$ relationship seen to be exhibited by most single O stars and WN and O star massive binaries. This does not narrow down our identification as many examples of overluminous ($\log(L_{x}/L_{bol}) > -7$) putatively single WN and O stars, as well as CWBs, with $\log(L_{x}/L_{bol}) \approx -6.5$, have been detected \citep[e.g.][]{oskinova05,sana06,antokhin08,mauerhan10}.

There is no evidence to suggest long term variability between the original \asca\ and recent \cxo\ observations of AX J144701--5919 and AX J144547--5931. This, combined with the lack of short term variability or periodicity down to 6.4s seen in the \cxo\ observations, is more supportive of a single star or CWB scenario rather than a SFXT scenario. However, longer X-ray observations are required to confirm this interpretation.

The radio spectral index of AX J144701--5919 is negative ($\alpha \approx -0.5$) and therefore dominated by non-thermal emission. This spectral index predicts a flux of $\sim38$ mJy at 843 MHz, which is much higher than the fluxes measured by MOST at this frequency ($\sim$10--15~mJy). This discrepancy could be due to time variability, or the radio spectrum may turn over at lower frequencies in a similar fashion to \object{WR 147} \citep{skinner99}. Further radio observations are required to distinguish between these two scenarios. Even though non-thermal radio emission supports a binary interpretation for the WN7--8h counterpart to AX J144701--5919, it does not allow us to discriminate between the possible origins of the X-ray emission.

In summary, the \cxo\ observations of AX J144701--5919 and AX J144547--5931 show these sources to have similar X-ray luminosities to those seen in putatively single massive stars, CWBs and quiescent SFXTs. As $\log(L_{x}/L_{bol}) > -6.5$  for both these sources, this supports both a binary and X-ray luminous isolated massive star scenario, the later of which is also supported by the fact that few hard X-ray photons ($kT>3$ keV) were detected. Longer X-ray observations to obtain statistically significant spectra are required to determine the mechanisms behind the X-ray generation from these two massive stars.

\section{Conclusions}

We have characterized the multi-wavelength properties of four previously unidentified \asca\ Galactic plane survey sources, AX J163252--4746, AX J184738--0156, AX J144701--5919 and AX J144547--5931, identifying them as X-ray emitting massive stars. Our data are most consistent with colliding-wind binary classifications for AX J163252--4746 and AX J184738--0156. AX J144547--5931 and AX J144701--5919 require longer X-ray observations to distinguish between a colliding-wind binary, a high-mass X-ray binary or an extremely X-ray luminous isolated massive star.

In the full \chic\ survey we have so far observed 95 \asca\ Galactic Plane Survey sources with \cxo, $\sim10\%$ of which emit hard X-rays (median energies $> 2$ keV) and have colors and infrared excess similar to the four objects discussed in this paper. One AGPS source with these characteristics, \object[AX J1831.2-1008]{AX J183116--1008}, has indeed recently been identified as a WN8 binary object, likely a colliding-wind binary, with an unabsorbed X-ray luminosity of $L_{x} \sim 1.3 \times 10^{34}$ erg s$^{-1}$ \citep{motch09pp}. It is possible that the other AGPS sources with similar X-ray and infrared properties to the four discussed in this paper could be more X-ray emitting CWBs or X-ray binary objects.

The counterparts to AX J163252--4746, AX J184738--0156, AX J144701--5919 and AX J144547--5931 as well as AX J183116--1008, are all extremely massive stars, making them short-lived and therefore rare, absorbed and only found in the Galactic plane. This makes the \asca\ Galactic Plane Survey and our follow-up \chic\ project, an efficient way of locating and identifying massive stars in our Galaxy. 

\begin{acknowledgements}

Special thanks goes to Michael Muno for his encouragement, expertise and participation in this project. We also thank Sean Farrell, Stan Owocki and Nathan Smith for their advice and help with this research, and the referee for their constructive response and suggestions. G.E.A acknowledges the support of an Australian Postgraduate Award. B.M.G. acknowledges the support of a Federation Fellowship from the Australian Research Council through grant FF0561298. D.L.K. was supported by NASA through Hubble Fellowship grant \#51230.01-A awarded by the STScI, which is operated by AURA, for NASA, under contract NAS 5-26555. P.O.S. acknowledges partial support from NASA Contract NAS8-03060. D.T.H.S. acknowledges a STFC Advanced Fellowship. J.J.D was supported by NASA contract NAS8-39073 to the Chandra X-ray Center (CXC). Basic research in radio astronomy at the NRL is supported by 6.1 Base funding. Support for this work was also provided by NASA through \cxo\ Award Number GO9-0155X issued by the CXC, which is operated by the Smithsonian Astrophysical Observatory for and on behalf of NASA. This research makes use of data obtained with the \cxo\ \textit{X-ray Observatory}, and software provided by the CXC in the application packages \texttt{CIAO} and \texttt{Sherpa}. OSIRIS is a collaborative project between Ohio State University and CTIO. Observing time on the 6.5m Clay and Baade Magellan Telescopes, located at Las Campanas Observatory, was allocated through the Harvard-Smithsonian Center for Astrophysics. The ATCA, part of the Australia Telescope, is funded by the Commonwealth of Australia for operation as a National Facility managed by CSIRO. This publication makes use of data products from the second catalogue of serendipitous X-ray sources (2XMMi) from the European Space Agency's (ESA) \textit{XMM-Newton} observatory. These data were accessed through the Leicester Database and Archive Service at Leicester University, UK. 2MASS is a joint project of the University of Massachusetts and the IPAC/Caltech, funded by the NASA and NFS. GLIMPSE survey data are part of the Spitzer Legacy Program. The \textit{Spitzer Space Telescope} is operated by the JPL/Caltech under a contract with NASA. This research has made use of NASA's Astrophysics Data System.

\end{acknowledgements}

{\it Facilities:} \facility{ASCA}, \facility{ATCA}, \facility{CXO (ACIS, HRC)}, \facility{Magellan:Baade (PANIC)}, \facility{Magellan:Clay (LDSS-3)}, \facility{Molonglo}, \facility{SOAR (OSIRIS)}, \facility{XMM (EPIC)}

\pagebreak


\pagebreak

\begin{turnpage}

\begin{deluxetable}{lccclcc}
\tablewidth{0pt}
\tabletypesize{\scriptsize}
\tablecaption{X-ray Observations of AX J163252--4746, AX J184738--0156, AX J144701--5919 and AX J144547--5931\label{tab:1}}
\tablehead{
\colhead{Source} & \colhead{X-ray Telescope} & \colhead{Obs ID} & \colhead{Date (yyyy-mm-dd)} & \colhead{Instrument} & \colhead{Exp Time (s)} & \colhead{Total Counts\tablenotemark{a}}
}
\startdata
AX J163252$-$4746 & \cxo & \dataset[ADS/Sa.CXO#obs/8155]{8155} & 2007-10-23 & HRC-I & 2670 & $68 \pm 8$
\\
 & \xmm & 0201700301 & 2004-08-19 & EPIC MOS1 & 50987 & $3539 \pm\ 65$
\\
 & & & & EPIC MOS2 & 51081 & $3304 \pm\ 62$
\\
 & & & & EPIC PN & 46742 & $7338 \pm\ 94$
\\
\\
AX J184738$-$0156 & \cxo & \dataset[ADS/Sa.CXO#obs/9612]{9612} & 2008-05-27 & ACIS-S & 1650 & $123 \pm\ 11$
\\
 & \xmm & 0203850101 & 2004-10-22 & EPIC MOS1 & 25131 & $2425 \pm\ 54$
\\
 & & & & EPIC MOS2 & 25155 & $2311 \pm\ 52$
\\
 & & & & EPIC PN & 20632 & $4939 \pm\ 77$
\\
\\
AX J144701$-$5919 & \cxo & \dataset[ADS/Sa.CXO#obs/8233]{8233} & 2007-01-13 & ACIS-S & 1450 & $83 \pm\ 9$
\\
\\
AX J144547$-$5931 & \cxo & \dataset[ADS/Sa.CXO#obs/8240]{8240} & 2007-08-12 & ACIS-S & 2550 & $69 \pm\ 8$
\enddata
\tablenotetext{a}{Total number of counts in the $0.3-8.0$ keV energy range for ACIS-S, the full \cxo\ energy range ($0.1-10.0$ keV) for HRC-I and $0.2-12.0$ keV energy range for \xmm.}
\end{deluxetable}

\begin{deluxetable}{lccccccccccccc}
\tablewidth{0pt}
\tabletypesize{\scriptsize}
\tablecaption{X-ray Positions and Spectral Modeling of AX J163252--4746, AX J184738--0156, AX J144701--5919 and AX J144547--5931\label{tab:2}}
\tablehead{
\colhead{Source} & \multicolumn{2}{c}{\cxo\ Position\tablenotemark{a}} & \colhead{} & \colhead{Telescope} & \colhead{} & \colhead{Distance} & \colhead{} & \multicolumn{5}{c}{Raymond-Smith Fit Parameters\tablenotemark{c}}
\\
\cline{2-3} \cline{5-5} \cline{7-7} \cline{9-13}
\\
\colhead{AX} & \colhead{RA(J2000)} & \colhead{Dec(J2000)} & \colhead{} & \colhead{Obs ID\tablenotemark{b}} & \colhead{} & \colhead{(kpc)} & \colhead{} & \colhead{$kT$ (keV)} & \colhead{$N_{H}$} & \colhead{Abundances\tablenotemark{d}} & \colhead{$F_{x,unabs}$} & \colhead{$\log(L_{x}/L_{bol})$\tablenotemark{e}}
\\
\colhead{} & \colhead{} & \colhead{} & \colhead{} & \colhead{} & \colhead{} & \colhead{} & \colhead{} & \colhead{$\chi^{2}_{red}$} & \colhead{$F_{x,abs}$} & \colhead{} & \colhead{$L_{x,unabs}$} & \colhead{}
}
\startdata
J163252--4746 & 16:32:48.548 & -47:45:06.20 & & \xmm & & $\sim4.5$ & & $3.2 \pm 0.2$ & $6.2 \pm 0.3$ & $0.9 \pm 0.1$ & $11.9^{+0.6}_{-0.5}$ & $\sim$ -5.4 -- -5.1
\\
& & & & 0201700301 & & & & 0.9 & $3.7 \pm 0.1$ & & $\sim3.4$ & 
\\
\\
J184738--0156 & 18:47:36.650 & -01:56:33.32 & & \xmm & & $\sim5.6$ & & $3.1 \pm 0.3$ & $8.9 \pm 0.5$ & $0.7 \pm 0.1$ & $9.1^{+0.9}_{-0.6}$ & $\sim$ -5.4 -- -5.0
\\
& & & & 0203850101 & & & & 0.8 & $2.2 \pm 0.1$ & & $\sim3.4$ & 
\\
\\
 &  &  & & \cxo & & & & 7.5 ($>1.9$) & $10.5^{+11.7}_{-5.0}$ & 0.7 &  & 
\\
& & & & 9612 & & & & 0.7 & 2.0 ($>0.3$) & & & 
\\
\\
J144701--5919 & 14:46:53.583 & -59:19:38.31 & & \cxo & & $\sim2.6$ & & $1.3^{+1.3}_{-0.5}$ & $2.9^{+1.7}_{-1.3}$ & 1.0 & $\sim4$ & $\sim$ -6.4 -- -6.0
\\
& & & & 8233 & & & & 0.2 & $0.5^{+1.0}_{-0.4}$ & & $\sim 0.32$ & 
\\
\\
J144547--5931 & 14:45:43.673 & -59:32:05.25 & & \cxo & & $\sim2.3$ -- 3.0 & & $1.2^{+8.9}_{-0.8}$ & $3.5^{+4.5}_{-2.1}$ & 1.0 & $\sim3$ & $\sim$ -6.5 -- -6.3 
\\
& & & & 8240 & & & & 0.3 & 0.3 ($<1.6$) & & $\sim$ 0.19 -- 0.32 & 
\enddata

\tablecomments{In the cases where one of the 90\% errors for a given spectral fit parameter was incalculable we quote the best fit parameter followed by a 90\% confidence upper or lower limit. The \xmm\ spectra were used rather than the \cxo\ spectrum to calculate the unabsorbed flux and luminosity of AX J184738--0156 as the model fits to the \xmm\ data are far more statistically significant.}

\tablenotetext{a}{All position errors are within a $0\farcs8$ radius circle for a 95\% confidence.}
\tablenotetext{b}{X-ray telescope and corresponding observation identification used in the spectral modeling.}
\tablenotetext{c}{The best fit Raymond-Smith model parameters including temperature $kT$ (keV), absorption column density $N_{H}$ ($10^{22}$ cm$^{-2}$), reduced chi-sqaure  $\chi_{red}^{2}$ from Chi Gehrels statistics and abundances relative to the solar value given by \citet{anders89}. The absorbed X-ray flux, $F_{x}$, is in the $0.3-8.0$ keV and $0.4-10.0$ keV energy range for the \cxo\ and \xmm\ fits, respectively. The unabsorbed X-ray flux, $F_{x,unab}$ ($10^{-12}$ \erg) and the unabsorbed X-ray luminosity, $L_{x,unab}$, ($10^{34}$ ergs s$^{-1}$) are calculated from the Raymond-Smith fits over the $0.3-8.0$ keV energy range. All fit parameter errors are for 90\% confidence.}
\tablenotetext{d}{The abundance parameter was prevented from varying during the  fitting process of the \cxo\ spectra. The parameter was set to the number listed, which is relative to the solar value given by \citet{anders89}. The abundance parameter was allowed to vary during the fitting of the \xmm\ spectra.}
\tablenotetext{e}{A range of possible $\log(L_{x}/L_{bol})$ calculated using the bolometric luminosity, $L_{bol}$ (ergs s$^{-1}$), of sources with similar spectral types taken from Table~3 of \citet{mauerhan10} and Table~5 of \citet{oskinova05}. This ratio was calculated using WN7--8h stars for AX J184738--0156 as its $K$-band spectral morphology \citep[see Figure~5 in][]{blum99} is similar to stars of this subtype seen in \citet{martins08} and \citet{mauerhan10}. In the case of AX J144547--5931 we used the value of $L_{bol}$ that \citet{mauerhan10} calculated for X-ray source \object[CXOGC J174617.0-285131]{CXOGC J174617.0--285131}, which is classified as an O6If$^{+}$ star.}

\end{deluxetable}

\begin{deluxetable}{lccccccccccccc}
\tablewidth{0pt}
\tabletypesize{\scriptsize}
\tablecaption{Infrared Magnitudes and Radio Fluxes of counterparts to AX J163252--4746, AX J184738--0156, AX J144701--5919 and AX J144547--5931 \label{tab:3}}
\tablehead{
\colhead{Source} & \multicolumn{3}{c}{2MASS Magnitudes\tablenotemark{a}} & \colhead{} & \multicolumn{4}{c}{GLIMPSE Magnitudes\tablenotemark{b}} & \colhead{} & \multicolumn{4}{c}{ATCA Fluxes (mJy)\tablenotemark{c}}
\\
\cline{2-4} \cline{6-9} \cline{11-14}
\\
\colhead{AX} & \colhead{$J$} & \colhead{$H$} & \colhead{$K_{s}$} & \colhead{} & \colhead{$m_{3.6}$} & \colhead{$m_{4.5}$} & \colhead{$m_{5.8}$} & \colhead{$m_{8.0}$} & \colhead{} & \colhead{1.4 GHz} & \colhead{2.4 GHz} & \colhead{4.8 GHz} & \colhead{8.0 GHz}
\\
\colhead{} & \multicolumn{3}{c}{Catalog Source Name} & \colhead{} & \multicolumn{4}{c}{Catalog Source Name} & \colhead{} & \colhead{} & \colhead{} & \colhead{} & \colhead{}
}
\startdata
J163252--4746 & 13.36 $\pm$ 0.04 & $>10.89$ & $>9.27$ & & 8.18 $\pm$ 0.06 & 7.43 $\pm$ 0.06 & 7.27 $\pm$ 0.03 & 6.76 $\pm$ 0.03 & & $<7.0$ & $<3.0$ & 2.6 $\pm$ 0.2 & 2.9 $\pm$ 0.3
\\
 & \multicolumn{3}{c}{2MASS J16324851--4745062} & \colhead{} & \multicolumn{4}{c}{SSTGLMC G336.5085+00.1571} & \colhead{} & \colhead{} & \colhead{} & \colhead{} & \colhead{}
\\
\\
J184738--0156 & 14.85 $\pm$ 0.07 & 11.01 $\pm$ 0.06 & 8.66 $\pm$ 0.04 & & & & 5.48 $\pm$ 0.19$^{*}$ & & & & & &
\\
& \multicolumn{3}{c}{2MASS~J18473666--0156334} & \colhead{} & \multicolumn{4}{c}{SSTGLMA~G030.7667--00.0346} & \colhead{} & \colhead{} & \colhead{} & \colhead{} & \colhead{}
\\
\\
J144701--5919 & 8.93 $\pm$ 0.02 & 7.60 $\pm$ 0.04 & 6.82 $\pm$ 0.02 & & 6.70 $\pm$ 0.09$^{*}$ & 6.33 $\pm$ 0.09$^{*}$ & 5.57 $\pm$ 0.03 & 5.32 $\pm$ 0.03 & & 27.0 $\pm$ 0.6 & 23.4 $\pm$ 0.5 & 16.8 $\pm$ 0.2 & 9.9 $\pm$ 0.2
\\
 & \multicolumn{3}{c}{2MASS J14465358--5919382} & \colhead{} & \multicolumn{4}{c}{SSTGLMC G317.1882+00.3106} & \colhead{} & \colhead{} & \colhead{} & \colhead{} & \colhead{}
\\
\\
J144547--5931 & 9.23 $\pm$ 0.04 & 8.10 $\pm$ 0.03 & 7.52 $\pm$ 0.02 & & 7.07 $\pm$ 0.05 & 6.89 $\pm$ 0.04 & 6.71 $\pm$ 0.03 & 6.63$\pm$ 0.02 & & & & &
\\
 & \multicolumn{3}{c}{2MASS J14454369--5932050} & \colhead{} & \multicolumn{4}{c}{SSTGLMC G316.9656+00.1864} & \colhead{} & \colhead{} & \colhead{} & \colhead{} & \colhead{}
\enddata

\tablecomments{AX J184738--0156 was detected in the GLIMPSE 3.6, 4.5 and 8.0 {\micron} bands but no magnitudes were obtained as the source was either saturated or the surrounding region too confused.}

\tablenotetext{a}{J, H and K$_{s}$ counterpart magnitudes are taken from the 2MASS All-Sky Catalog of Point Sources \citep{skrutskie06}. Due to blending with nearby sources only 95\% confidence lower limits are quoted for the $H$ and $K$ magnitudes of AX J163252--4746.}
\tablenotetext{b}{$m_{3.6}-m_{8.0}$ refers to the counterpart magnitudes at $3.6-8.0$ {\micron} taken from the GLIMPSE {\sc i} DR5 Catalog. Magnitudes labeled with $^{*}$ are from the GLIMPSE {\sc i} DR5 Archive, which is described as more complete but less reliable than the GLIMPSE {\sc i} DR5 Catalog.}
\tablenotetext{c}{ATCA radio observations were only obtained for AX J144701--5919 and AX J163252--4746. The $1\sigma$ flux errors are quoted. No radio counterpart to AX J163252--4746 was detected at 1.4 and 2.4 GHz so 5$\sigma$ flux upper-limits are quoted.} 

\end{deluxetable}

\end{turnpage}

\pagebreak

\end{document}